\documentclass[12pt]{iopart}

\usepackage{iopams}
\usepackage[us,12hr]{datetime}
\usepackage{graphicx}

\usepackage[dvips, bookmarks, colorlinks=true, pdftitle={How cold can you get in space? Quantum Physics at cryogenic temperatures in space}, pdfsubject={How cold can you get in space?}, pdfkeywords={Space mission, Cryogenic instrument,
Quantum mechanics, Macrorealism, Quantum decoherence, Quantum optomechanics}]{hyperref}
\begin{document}
\title[How cold can you get in space?]{How cold can you get in space? Quantum Physics at cryogenic temperatures in space}
\author{Gerald Hechenblaikner$^1$, Fabian Hufgard$^1$, Johannes Burkhardt$^1$, Nikolai Kiesel$^2$, Ulrich Johann$^1$, Markus Aspelmeyer$^2$, and Rainer Kaltenbaek$^2$}
\address{$^1$ EADS Astrium, 88039 Friedrichshafen, Germany}
\address{$^2$ Vienna Center for Quantum Science and Technology, Faculty of Physics, University of Vienna,
Vienna, Austria}
\begin{abstract}
Although it is often believed that the coldness of space is ideally suited for performing measurements at cryogenic temperatures, this must be regarded with caution for two reasons: Firstly, the sensitive instrument must be completely shielded from the strong solar radiation and therefore, e.g. either be placed inside a satellite or externally on the satellite's shaded side. Secondly, any platform hosting such an experiment in space generally provides an environment close to room temperature for the accommodated equipment. To obtain cryogenic temperatures without active cooling, one must isolate the instrument from radiative and conductive heat exchange with the platform as well as possible. We investigate the limits of this passive cooling method in the context of a recently proposed experiment to observe the decoherence of quantum superpositions of massive objects. The analyses and conclusions are applicable to a host of similar experimental designs requiring a cryogenic environment in space.
\end{abstract}

\vspace{2pc}

\maketitle
\section{Introduction}
Experiments often require isolating the object under
investigation from its environment. This holds true in particular for quantum experiments, where any information shared
with the environment
may decohere the quantum state and disturb
its evolution in time. Free and undisturbed evolution in time is, however, an essential prerequisite in many
quantum experiments, and it is even more important, when the laws of quantum
physics themselves shall be put to the test.

Consider a physical system left alone in outer space far from any other objects. This is probably the situation closest to absolute isolation we can imagine. Would we be able to completely describe the evolution of the system in terms of quantum physics? Maybe not. For example, spatial superpositions of massive objects
might behave unexpectedly due to the yet unclear role of gravity in the context of quantum physics. A number
of modifications to quantum theory have been suggested
that predict decoherence of massive quantum superpositions even for completely isolated systems. Among such proposed theoretical extensions of standard quantum theory are the ``macrorealistic'' models of Di\'osi\cite{diosi1984gravitation}, Penrose\cite{penrose1996gravity},
K\'arolyh\'azy\cite{karolyhazy1966gravitation}, the continuous-spontaneous-localisation (CSL) model \cite{pearle1976reduction,ghirardi1986unified,gisin1989stochastic}, and the quantum-gravity model of Ellis\cite{ellis1984search}. A detailed overview of such models and of experiments testing them is given in Refs.\cite{RomeroIsart2011c,bassi2013models}. Examples of Earth-based experiments towards realizing macroscopic quantum superpositions are Refs.~\cite{brune1996,Fickler2012a} using photon states, Refs.~\cite{Friedman2000a,VanderWal2000a} using superconducting loops, Ref.~\cite{Julsgaard2001a} using spin-squeezed atomic ensembles, and Refs.~\cite{Arndt1999C60,Gerlich2011a} using molecules made up from a large number of atoms. 

Naturally, no experiment will be able to realize the idealized situation described above -- first and foremost, because a completely isolated particle can
neither be prepared nor measured and thus cannot be used to test our
predictions. However, we can ask how close we can get to this situation. Whereas earth-bound experiments have the natural limitation that free-fall experiments cannot be continued over very long times, this limitation is lifted when going to space. In this paper, we analyse design provisions for optimal thermal isolation of an experimental platform accommodated externally to a spacecraft. Specifically, we focus on the thermal isolation of a non-tangible, evacuated test volume surrounding a massive test particle from the hot spacecraft surface by appropriate thermal shielding. The corresponding optimization of the shield design is performed for the main instrument of the recently proposed MAQRO ``Macroscopic quantum resonators''\cite{kaltenbaek2012macroscopic} mission that aims to test the validity of the quantum superposition principle for massive objects against modifications to quantum theory as mentioned above.

In order to derive design constraints of the thermal shield that are custom-tailored for the proposed quantum-decoherence experiments, we start with a brief
summary of the MAQRO mission. We then proceed to give
a detailed account of the optimization procedure for the thermal shielding
that protects the experimental platform. The optimization is based on
simulations that can easily be adapted to other experiments and
platforms. Finally we demonstrate that the proposed design of the
radiation shield indeed provides sufficient isolation of the experiment to
perform meaningful tests at the foundations of quantum theory. Our analyses and the final design are suitable as a reference for similar science experiments in space.

\section{Macroscopic Quantum Oscillators in Space}
\subsection{Mission design}
The proposed MAQRO \cite{kaltenbaek2012macroscopic} space mission aims to explore quantum physics in yet untested parameter regimes by observing the decoherence of superpositions of macroscopic objects.
Isolating the quantum system from the environment as well as possible is essential for MAQRO. This requires:
\begin{itemize}
\item a low internal temperature of the quantum system to minimize decoherence due to the emission of black-body radiation.
\item a low environment temperature to minimize decoherence due to the absorption and scattering of black-body radiation.
\item ultra-high vacuum to prevent scattering of gas molecules by the quantum system.
\end{itemize}
Here, we will mainly consider the second requirement. In our design, using radiation shields for passive cooling, direct outgassing into space at low environment temperatures automatically leads to the fulfilment of the ultra-high-vacuum requirement. Achieving a low environment temperature is, therefore, a key requirement for the mission design. The latter has already been described in detail in Ref.~\cite{kaltenbaek2012macroscopic}. Here, we shall only briefly review some relevant features. While some of the presented design features may be specific to MAQRO, the subsequent analyses for the thermal shield are set in a broader frame, rendering our results applicable to general designs of instruments requiring very low temperatures.

In the past years, a number of science missions have been developed where the experimental apparatus is cooled using a reservoir of liquid Helium \cite{Everitt2011GPB}, \cite{pilbratt2008herschel}. While this allows reaching cryogenic temperatures, it comes at the expense of high cost and complexity as well as a lifetime limited by the depletion of coolant. In other missions, the coldness of space (roughly $3\,$K background temperature) is exploited in passive cooling concepts where the instrument faces deep-space behind a multi-layer radiation shield protecting it against solar radiation in a sun-synchronous orbit. An application example is a spectrometer recording images with as little thermal disturbance as possible. One such sun-synchronous orbit, a halo orbit around the L2-Lagrange point of the earth-sun system, was chosen for the James-Webb space telescope\cite{Lig2012}, the Herschel/Planck and Gaia missions\cite{hechler2002herschel}, and is similarly proposed for MAQRO.
This type of orbit is minimally afflicted by external perturbations and allows keeping the spacecraft stably pointed towards the sun throughout the mission. In addition to simplifying the power and thermal architecture of the satellite, this offers optimal experimental conditions: On the one hand, simple body-mounted solar arrays can be used without need for a solar-array driving mechanism. On the other hand, excellent temperature stability is inherently provided as a result of the uniform incidence of solar flux.
\begin{figure}
\begin{center}
\begin{tabular}{l}
\includegraphics[width=14cm]{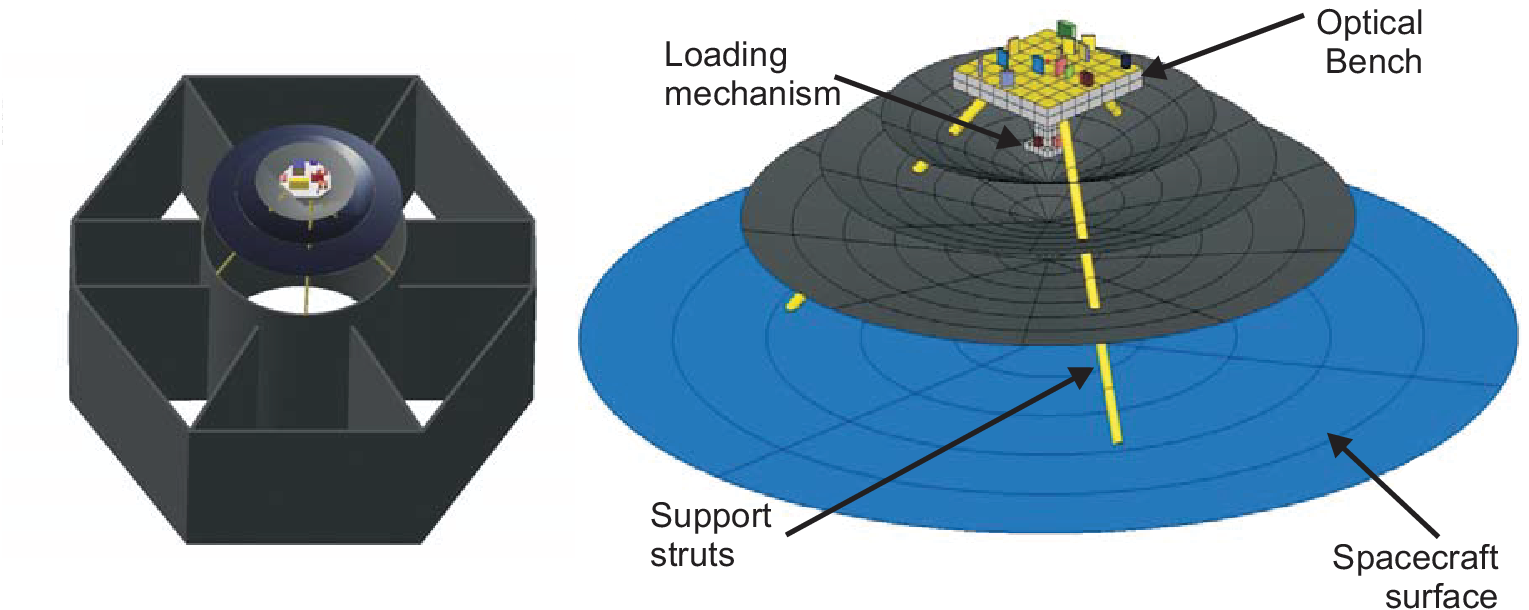}

\end{tabular}
\end{center}
\caption[example]

{ \label{fig:DECIDE_total}
(Left) CAD-model of the MAQRO instrument attached to the structural cylinder of a typical spacecraft. (Right) We used a geometric surface model in the simulations for the thermal analysis.}
\end{figure}
The central component of the MAQRO instrument is an optical bench that is accommodated externally on the shaded side of the
satellite and shielded from the ``hot'' spacecraft surface by several layers of radiation shields. The ``warm'' electronic units of the instrument, except for the sensor, are all accommodated in the spacecraft. This architecture is illustrated schematically in Fig. \ref{fig:DECIDE_total} (left) using the LISA Pathfinder spacecraft\cite{mcnamara2008lisa} as a reference for an L1/L2 platform. The bench is surrounded by the innermost shield and mounted on supporting struts that are fixed to the spacecraft inner structural cylinder. Choosing the shield dimensions for the assembly to fit into the structural cylinder simplifies instrument accommodation and provides the possibility of using extensions of the cylinder as protective enclosure before the extension is discarded during commissioning.

\subsection{Experimental Setup}
MAQRO aims at exploring the quantum-mechanical concept of superposition for massive particles. To this end, a dielectric nanosphere with a radius between $90\,$nm and $120\,$nm and a mass of $\sim 10^{10}\,$amu is loaded from a dispensing mechanism (see Fig. \ref{fig:DECIDE_total} right) into an optical trap. The trap is formed by a Gaussian cavity mode. Once the particle is trapped, it is cooled close to the quantum mechanical ground state by a combination of cavity cooling\cite{horak1997cavity,barker2010doppler,kiesel2013cavity} and feedback-cooling \cite{gieseler2012subkelvin,li2011millikelvin}. After cooling the particle is released from the trap by switching off the optical fields. The wavefunction will then expand freely for a time on the order of $1\,$s. After that time, the particle is prepared in a superposition state of two positions by the action of a weak UV-pulse\cite{kaltenbaek2013testing} or by using cavity-optomechanical interactions\cite{romero2011large}. Then the superposition state is allowed to expand freely for another period of time on the order of $100\,$s. This is necessary for the two parts of the superposition state to overlap and form an interference pattern. In order to measure this interference pattern, the optical fields are switched on again, and the particle position along the cavity axis is measured via a combination of scattered-light imaging and cavity readout.
This procedure is repeated many times over to reconstruct the interference pattern and to determine the interference visibility. From the latter, one can determine the decoherence rates.

\begin{figure}
\begin{center}
\begin{tabular}{c}
\includegraphics[width=14cm]{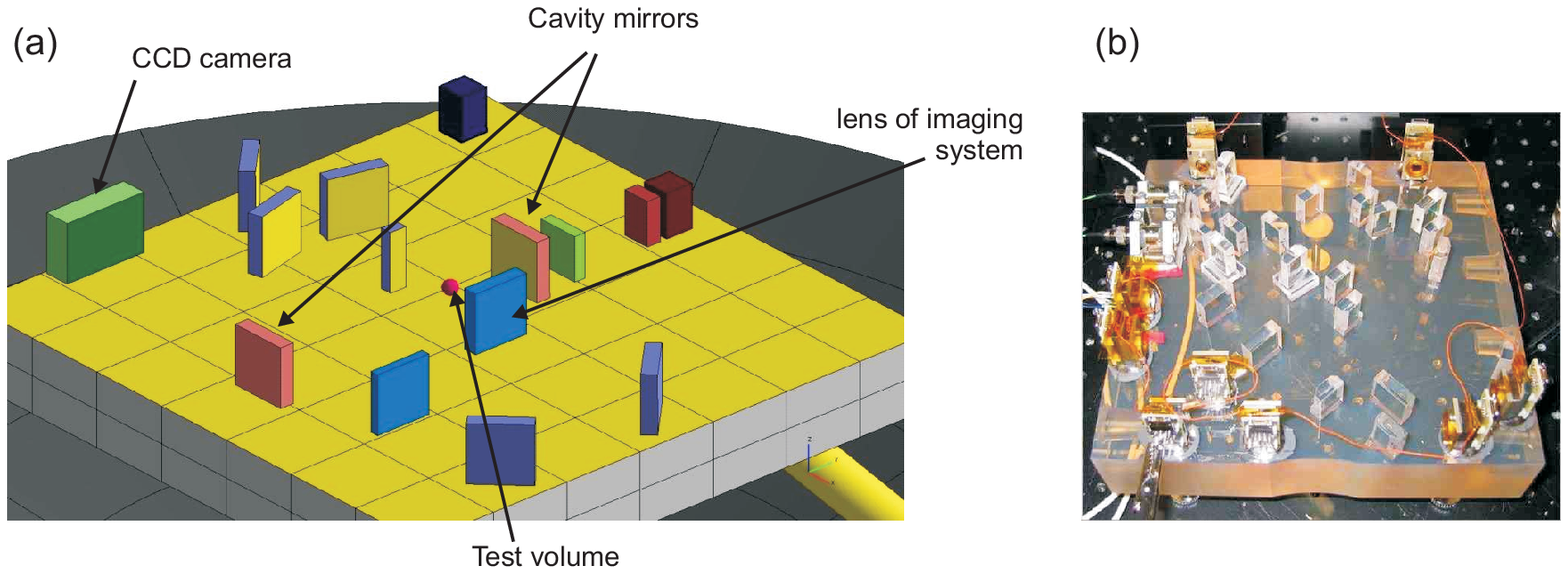}

\end{tabular}
\end{center}
\caption[example]

{ \label{fig:optical_bench}
(Left) Close-up of the optical bench from Fig.\ref{fig:DECIDE_total}. Its base-plate measures 20 cm x 20 cm x 2 cm. (Right) The optical bench of LISA Pathfinder for comparison\cite{mcnamara2008lisa}.}
\end{figure}

A simplified representation of the layout of the optical bench (as used for thermal modelling) is presented in Fig.\ref{fig:optical_bench} left. The optical bench is proposed to be built from components made of silicon carbide (SiC), Zerodur and fused silica. SiC is a material with a very low coefficient of thermal expansion (CTE) of significantly less than $10^{-7}~{\rm K^{-1}}$ at very low temperatures. This type of material has also been used in the near-infrared spectrograph (NIRSpec) of the James-Webb space telescope\cite{Honnen2008nirspec} and in the instrument module of the Gaia mission \cite{bougoin2011herschel}. Zerodur exhibits a very low CTE at room temperature and has been used for the optical bench of LISA Pathfinder (Fig.\ref{fig:optical_bench} right), where hydroxide-catalysis bonding of optical elements was successfully applied to obtain a quasi-monolithic structure of superb stability\cite{elliffe2005hydroxide}. Despite the superior material properties of SiC at very low temperatures $< 30\,$K, i.e. the regime we aim for as discussed in section \ref{sec::analysis}, our current model assumption for the optical bench substrate is Zerodur, which facilitates manufacture and allows using a qualified bonding process.

\section{Thermal analysis for a cryogenic instrument}
\label{sec::analysis}
In the proposed instrument design, three conical thermal shields surround the optical bench in a concentric arrangement. They shield it from radiative heat exchange with the ``hot'' exterior surface of the spacecraft (see Fig.\ref{fig:DECIDE_total}). While the spacecraft interior is typically kept at room temperature ($\sim$300 K), as required for equipment operation, the external temperature of the shaded panel may drop as low as 120 K by effective use of multi-layer-insulation (MLI) sheets on the surface of the spacecraft. Motivated by comparison with other missions featuring primarily passively cooled cryogenic instruments, e.g. the James Webb space telescope \cite{Lig2012}, we shall aim for a somewhat lower and therefore more ambitious target temperature of $30\,$K for our optical bench. This temperature is limited by a combination of three effects:
\begin{itemize}
\item Radiative heat exchange by emission of thermal photons
\item Conductive heat exchange through material-components (e.g. struts and wires)
\item Electrical and optical dissipation on the optical bench
\end{itemize}
The thermal model of the instrument was constructed from more than 1000 thermal nodes and implemented in ESATAN-TMS software\cite{ESATAN2010}, a standard European thermal analysis tool for space systems. A large amount of nodes was required to model details of the optical bench whereas a coarser grid was used for the spacecraft surface and the shields. All nodes of the model can be coupled radiatively and conductively to any of the surrounding nodes. In order to include external influences in the model, e.g. the temperature of the spacecraft interior, we could use boundary nodes set to specific temperatures. Similarly, dissipation values could be assigned to individual nodes to model, e.g. electrical and optical dissipation in CCD head and cavity mirrors, respectively.

The radiative coupling parameters $GR_{ij}$ between surface elements $i$ and $j$ can be determined from the emissivities $\epsilon_i,\epsilon_j$ and the view factors $F_{ij}$ of the respective surface elements $i$ and $j$ via the relation $GR_{ij}=\epsilon_i\epsilon_j A_i F_{ij}$, where $A_i$ is the area of surface element $i$. The view factors are determined from a geometric surface model of the instrument, where the proportion of thermal flux emitted by one surface element and received by another is found for each pair of surface elements through Monte Carlo simulations.

The conductive coupling parameters $GL_{ij}$ depend on the thermal conductivity $\kappa$ of the material, the interface area $A$ between two segments, and the distance $d_{ij}$ between nodes. This dependence can be expressed via the relation: $GL_{ij}=\kappa\cdot A/d_{ij}$. The thermal conductivities at very low temperature were obtained from existing data sets for the Gaia and Herschel missions and the corresponding tables were included in the analyses\cite{hufgard2013thermal}.

\subsection{Radiative energy exchange}
\begin{figure}
\begin{center}
\begin{tabular}{c}
\includegraphics[width=14cm]{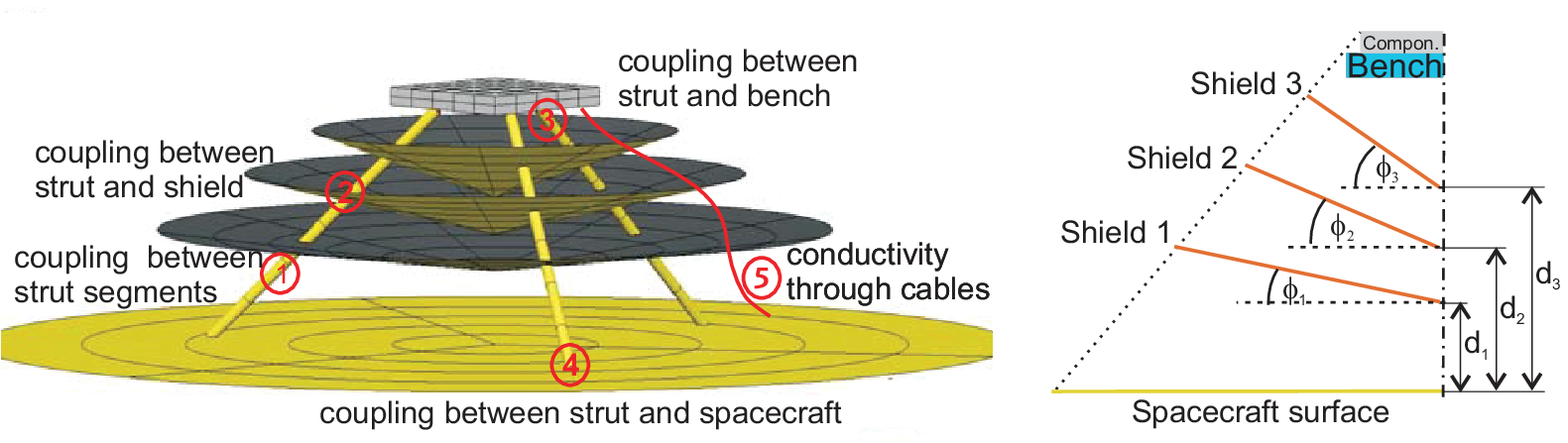}

\end{tabular}
\end{center}
\caption[example]

{ \label{fig:thermal_analysis_setup}
(Left) The basic processes for thermal conduction. (Right) The geometry parameters of the radiation shields. }
\end{figure}

In a first step, we aimed at optimizing the number and geometry of the radiation shields while heat conduction and dissipation were neglected. A graphical representation of the shields is shown in Fig.\ref{fig:thermal_analysis_setup} (right), where the dotted line demonstrates that neither the optical bench nor any of its components are in direct field of view with any part of the spacecraft surface, thereby blocking any direct exchange of thermal photons. The main idea behind the geometric design is that the shields are fanned by successively increasing their opening angle $\phi_i$ to space. Through this method, the radiative coupling between the two outer shields and the cold void of space is improved with respect to a plane-parallel geometry. The coupling to space is further stimulated by covering the upper side of the shields with a highly emissive material (black finish), while impairing coupling in between shields by covering their underside with a low-emissivity material (gold finish). Care must be taken that the opening angle of the inner shield $\phi_3$ is not too large as this would increase the radiative coupling to the optical bench with a corresponding increase of its temperature. Therefore, the optimum geometry must strike a fine balance between all these effects while also considering an adequate distance between spacecraft and shields.

\begin{figure}
\begin{center}
\begin{tabular}{c}
\includegraphics[width=14cm]{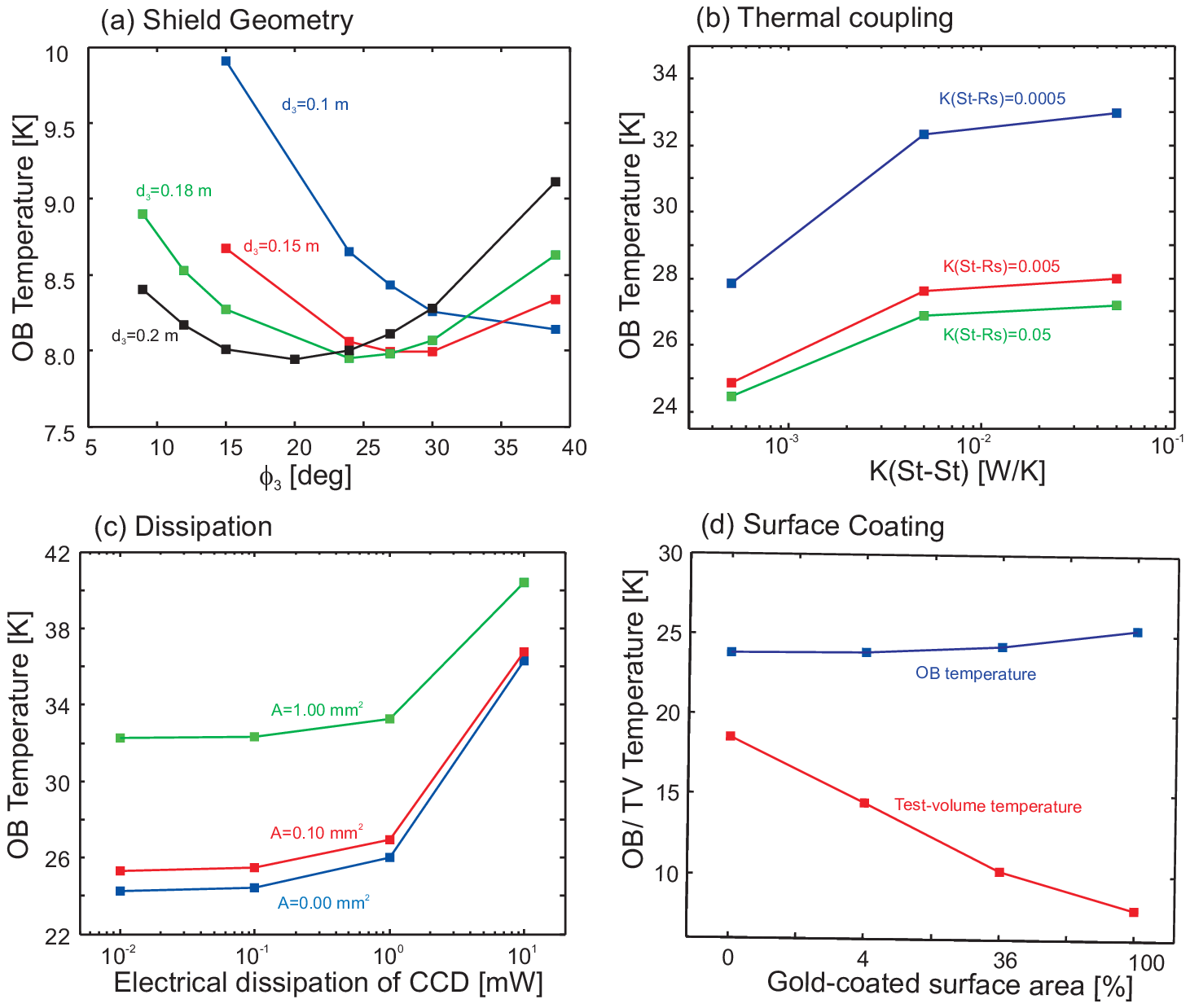}

\end{tabular}
\end{center}
\caption[example]

{ \label{fig:shield_geometry}
(a) The optical-bench temperature is plotted against the opening angle of the inner shield $\phi_3$ for various distances to the spacecraft $d_3$. (b) The optical-bench temperature is plotted against the thermal coupling parameter between strut segments $GL_{st,st}$ for various values of the coupling parameter between strut segments and radiation shields $GL_{st,rs}$. (c) The optical-bench temperature is plotted against the value of electrical dissipation of the CCD head for various values of the electrical-harness cross-section $A$. The simulations included an optical dissipation of $0.2\,$mW. (d) The temperatures of the optical bench and test-volume are plotted against the size of the gold-coated top surface area of the optical bench. The gold-coated area is centred around the test volume and the coated area's size is expressed as a percentage of the total top surface area.}
\end{figure}

For our analyses, the spacecraft surface was modelled with a circular shape of $1.4\,$m diameter (see blue area in Fig.\ref{fig:DECIDE_total} right) and the optical bench was kept at a fixed distance of $32.5\,$cm from that surface. This approach allows for a low-mass and compact shield design of a diameter only slightly larger than the structural cylinder of the spacecraft as shown in Fig.\ref{fig:DECIDE_total} (left).
While, in principle, all three shield opening angles and distances to the spacecraft can be optimized, at first only the geometric parameters of the inner shield, $\phi_3$ and $d_3$, were varied. The geometric parameters of the other shields were obtained via equipartition of the inner-shield parameters through the following relations: $\phi_1$=$1/3~\phi_3$, $\phi_2$=$2/3~\phi_3$, and $d_1$=$1/2~d_3$, $d_2$=$3/4~d_3$. Note that these constraints were confirmed to be close to optimal in subsequent analyses\cite{hufgard2013thermal}.
The results of the optimization are plotted in Fig.\ref{fig:shield_geometry}a.
The data show that the ideal opening angle $\phi_3$ varies between 15 and 30 deg depending on the distance of the shield to the spacecraft $d_3$. From this and similar analyses, an optimum temperature of $T_{\rm min}$=$8\,$K is found for an opening angle $\phi_3$= 20 deg at a distance of 20 cm. We also determined that, if only 2 shields are used instead of 3, the minimum temperature rises to 15 K, which is a rather large increase and is deemed unacceptable. On the other hand, adding another solid metal shield only yields a small further reduction in temperature. The corresponding performance gain seems unjustified when traded against the higher cost and complexity. A simpler alternative is to add additional shield layers in the form of MLI sheets, which reduces the temperature by 2 to 3 K. This is discussed in the next section \ref{subsec::conductivity}. For these reasons, we fix the design to a number of three shields and the geometric values specified above and proceed to the next step in the analysis.

\subsection{Thermal conductivity}
\label{subsec::conductivity}
Heat conduction through any material that connects to the bench, including mechanical support struts, electrical wires and optical fibres, constitutes the biggest challenge in achieving cryogenic temperatures. For that purpose, we took utmost care to base the design on materials with low conductivity and to minimize the conduction across critical material-joints. Various ways of conductive heat transfer and the corresponding couplings are depicted in Fig.\ref{fig:thermal_analysis_setup} left.

Central to our design concept are the 3 mechanical struts which are composed from 4 segments of glass-fibre-reinforced polymer that are joined by titanium end fittings at the penetration point of each shield. This allows obtaining low coupling parameters $GL_{st,st}$ between strut segments. While it is essential to minimize the coupling between the strut segments to increase the thermal resistance of the heat flow, it turns out to be advantageous to maximize the coupling between the struts and the radiation shields, described by the parameter $GL_{st,rs}$. This can be explained by the cooling capacity of the shields, which remove heat from the struts and thereby successively reduce the amount of heat transported to the optical bench. This relationship becomes apparent from Fig.\ref{fig:thermal_analysis_setup}b, where an optical-bench temperature of $T_{ob}\sim$ 27 K is obtained for realistic coupling parameters around $0.05\,$W/K.

These results indicate the significance of solid metal shields for helping to cool the support struts in addition to their primary role as radiation shields. In a renewed effort to improve the radiation shielding beyond the efficiency of the 3-layer solid metal design adopted so far, 3 more layers were added as simple MLI sheets. These were affixed to the solid metal shields on top of spacers in the computer model, which yielded a further reduction in the optical-bench temperature by 2-3 K. Another coupling parameter, $GL_{st,ob}$, which describes the coupling between the struts and the optical bench, seemed to be less important for the optical-bench temperature. Reducing this parameter by 2 orders of magnitude only decreased the temperature by 1 K.

Aside from the mechanical-support struts, the second major medium for heat conduction is the electrical harness (made from low-conductive steel) which connects to the CCD head on the optical bench. As shown in Fig.\ref{fig:shield_geometry}c, the effect on the optical bench temperature is only moderate as long as the wire cross section does not exceed $0.1~{\rm mm^2}$. The heat transfer through the optical fibres is relatively small and can be neglected in comparison to the electrical wires.

\subsection{Dissipation}
Dissipation is the final contribution to be included in the thermal analysis. The simulation results shown in Fig.\ref{fig:shield_geometry}c were performed with a detailed model of the optical bench (see Fig.\ref{fig:optical_bench}). The model included $0.2\,$mW of optical dissipation in addition to the electrical dissipation of the CCD head for various harness cross-sections $A$. The plots demonstrate how the temperature rapidly increases once the electrical dissipation exceeds $1\,$mW. Fortunately, dissipation as low as $1\,$mW constitutes a realistic design goal which can be achieved with a state-of-the-art CCD chip like the one used in the MIRI instrument of the James-Webb space telescope \cite{Love2004}. An additional electrical dissipation of 10 mW in a pre-processing chip, which was used in the design of all cryogenic JWST instruments \cite{loose2005sidecar}, was included in our thermal analyses. We found it to be uncritical once we placed the chip below the outer-shield, well within a reasonable distance of $<0.5\,$m between chip and CCD head.

\begin{figure}
\begin{center}
\begin{tabular}{c}
\includegraphics[width=12cm]{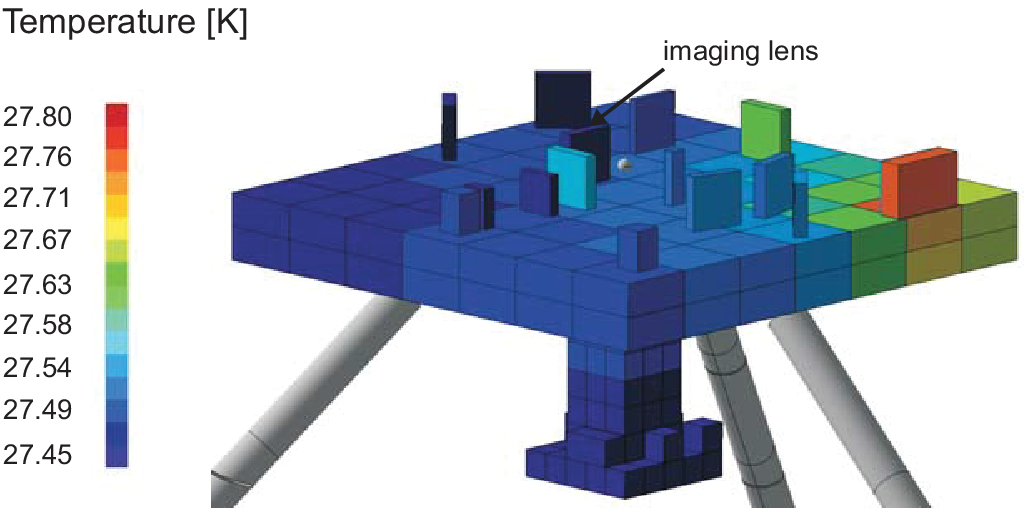}

\end{tabular}
\end{center}
\caption[example]

{ \label{fig:surface_temperature} The temperature of the gold-coated optical bench including all radiative, conductive, and dissipative effects. The dissipation of the CCD head (1 mW) and of the cavity mirrors ($<0.2\,$mW) affects an increase in temperature of the respective components.
}
\end{figure}

In summary, after choosing optimal but still realistic values for the thermal coupling parameters and electrical dissipation, we obtain an optical-bench temperature of $\sim 27\,$K. This defines the ultimate limit we can reach, based on quite generic design assumptions for radiation shields, mechanical structures, conductive heat transfer and dissipation without using active cooling. Fig.\ref{fig:surface_temperature} shows the temperature of the thermal nodes defining the optical bench, where the bench surface was modeled with an applied gold-coating (see section \ref{ExVol}). Due to the good thermal conductivity of the optical bench, the temperature varies only slightly across the bench. The temperature peaks at the CCD head and at the two cavity mirrors, where most of the electrical and optical power are dissipated, respectively.

\section{The Experiment}
\subsection{\label{ExVol} Driving Factors}
So far, we have aimed to optimize the experimental design with respect to the temperature of the optical bench. However, what really counts in the experiment (from a thermal perspective) is the effective temperature of the test volume located above the bench where the macroscopic quantum superposition evolves in free fall.

The temperature of the test volume is determined by the photon flux received by it and can be reduced by decreasing the photon emission into the volume. This can be achieved by coating the surface of the optical bench beneath the test volume with a material of low emissivity, such as gold. Figure \ref{fig:shield_geometry}d shows the simulation results for optical bench and test-volume temperatures when the area of surface coating is increased from zero to hundred percent. While this leads to only a moderate increase in the optical bench temperature due to the reduced coupling to space, the temperature of the test volume is dramatically decreased from 18 K to 8 K. These results clearly demonstrate the benefits of a low-emissivity surface coating for reducing the photon flux which by far outweighs the reduced cooling efficiency. For this reason, the optical bench is chosen to have a low-emissivity gold coating in the final design.

Note that, to facilitate the simulations underlying the data shown in Fig. \ref{fig:shield_geometry}d, all the components of the optical bench as well as the electrical harness were removed, leaving a plain bench surface without any protruding elements. After placing these components back onto the gold-coated optical bench, another thermal analysis was performed to yield the results for the final configuration. It was somewhat surprising to find that the temperature of the experimental volume jumped from roughly 8 K for the naked bench to 16 K for the populated one. After repeating the process of re-populating the bench, this time adding one component after the other and performing a thermal analysis after each step, we found that a single critical item is responsible for almost the full rise in temperature: the collecting lens of the imaging system. Whereas other optical components, like the cavity mirrors, are quite far from the test volume and covered by a gold-coating to minimize thermal emission, the uncoated imaging lens is in close proximity to the test volume and highly emissive.
Consequently, the photon emission from the imaging lens constitutes the limiting factor for the temperature of the test volume. That temperature was found to be $16.4\,$K after taking into account all conductive and dissipative effects and the final material properties for emissivity and thermal coupling. The temperature of the test volume may be reduced some more if the lens is placed further away from it or the lens diameter is decreased. However, this comes at the expense of a reduced numerical aperture of the imaging system. Therefore, the benefit of even lower temperatures must be carefully balanced against the penalty of a reduced resolution in any modification of the bench design. If the lens is removed, this reduces the temperature of the test volume to approximately $12\,$K, which gives an indication of the possible improvement for an optimized optical setup.

\subsection{Simulated experimental results}
Now that the instrument design has been optimized to obtain the lowest-possible temperatures in the experimental volume above the optical bench, we shall investigate the corresponding implications for the experimental measurements. In particular, we will consider the implications for distinguishing the predictions of quantum theory and the predictions of various macrorealistic models. As discussed in the introduction, keeping the temperature of the experimental volume as well as the internal temperature of the nanosphere very low, is a key requirement for a successful experiment. Other essential requirements, such as very low background pressure, very low levels of acceleration noise, and very long free-fall times have already been shown to be attainable in space\cite{kaltenbaek2013testing}. In fact, these requirements present the primary motivation for going into space because such experimental conditions cannot be achieved on ground.
Our proposal for a passively cooled instrument without using liquid Helium also avoids contamination of the experimental volume with Helium molecules. These are highly diffusive and present a serious problem: any collision of a Helium molecule with a the nanosphere would lead to a localisation of the quantum state and must be avoided by all means.

\begin{figure}
\begin{center}
\begin{tabular}{c}
\includegraphics[width=0.45\linewidth]{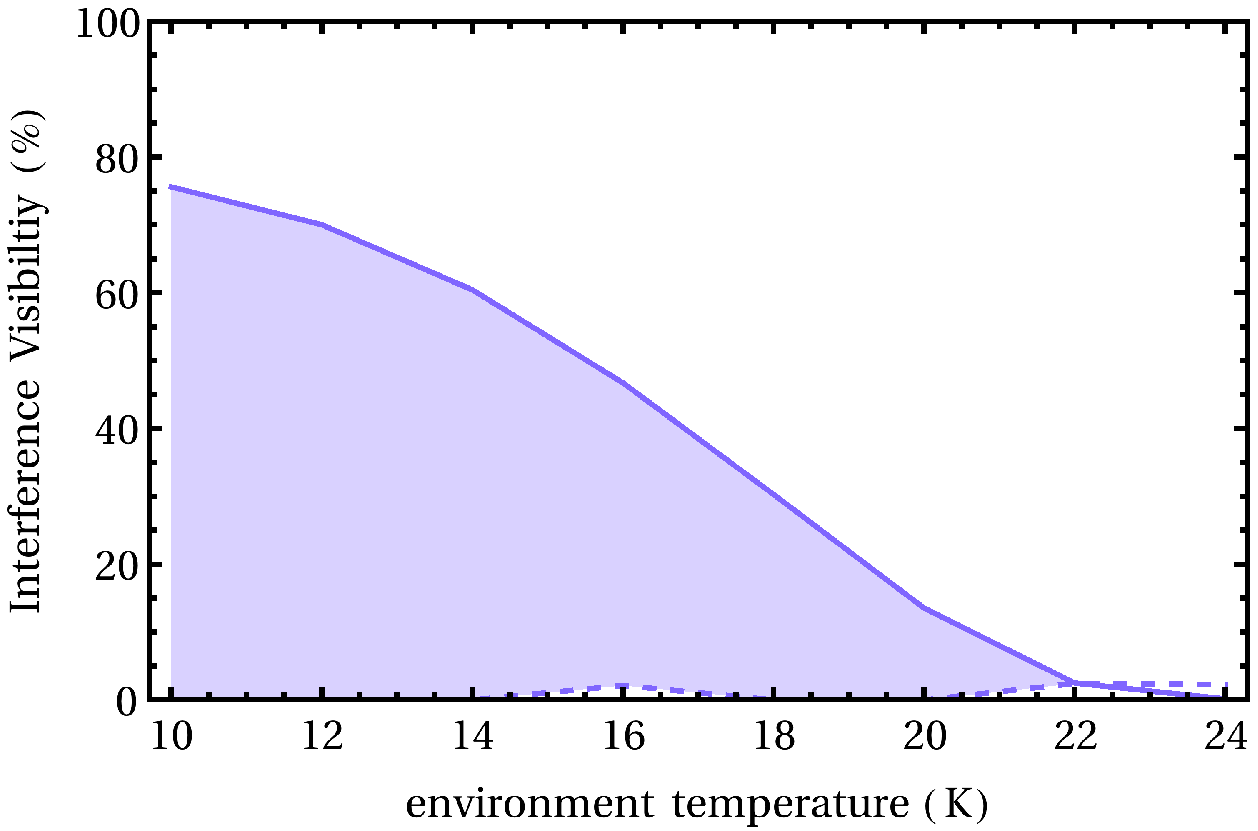}
\includegraphics[width=0.45\linewidth]{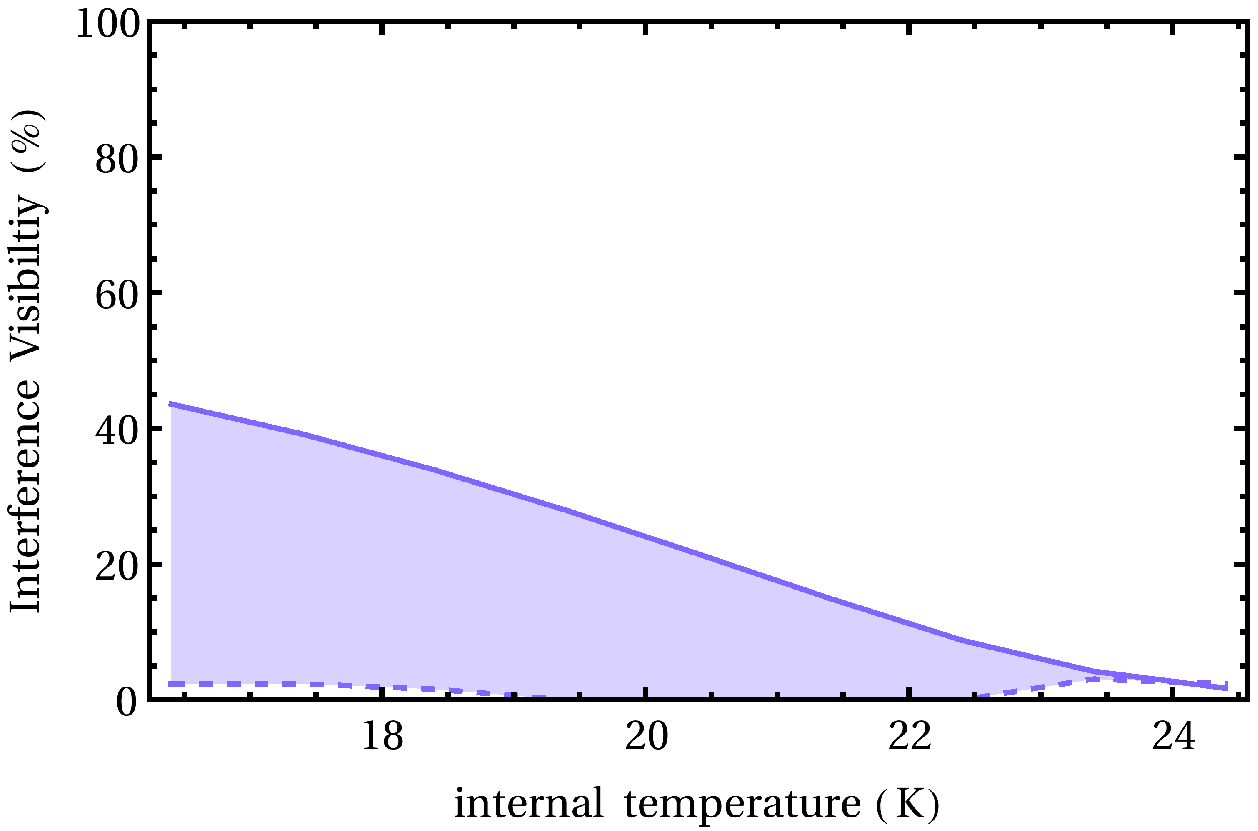}

\end{tabular}
\end{center}
\caption[example]

{ \label{fig:collapse_models} The expected interference visibility is plotted against environmental (left) and internal (right) temperature of the nanosphere, assuming an environmental temperature of $16.4\,$K for the right plot. The solid lines correspond to the predictions of quantum theory, the dashed lines represent the predictions of the macrorealistic model of K\'arolyh\'azy. Its prediction is zero apart from small numerical uncertainties in the numerical simulation. All other macrorealistic models predict zero visibility for the given parameter regime. Therefore, a test of all models is possible for a temperature where the quantum-theoretical prediction for the visibility exceeds the highest prediction of all alternative models (shaded region).
}
\end{figure}

The technical requirements of MAQRO are chosen such that they allow, in principle, to test most macrorealistic extensions of quantum theory known today. In particular, if we observe a non-zero interference visibility for nanospheres with a mass of $\sim 10^{10}\,$amu, this would already rule out the quantum-gravity model of Ellis\cite{ellis1984search}, and it would largely rule out the CSL model -- at least over a vast parameter regime including the original parameters proposed in Refs.~\cite{ghirardi1986unified,pearle1976reduction}. Moreover, MAQRO would allow for testing the models of Di\'{o}si\cite{diosi1984gravitation} and Penrose\cite{penrose1996gravity}, and it may even allow for testing the model of K\'arolyh\'azy\cite{karolyhazy1966gravitation}. Of course, this depends on how well we can isolate our quantum system from the environment. In order to test macrorealistic extensions of quantum theory, quantum theory itself has to predict a non-vanishing interference visibility. If we assume that the vacuum is good enough ($\lesssim 10^{-13}\,$Pa) to allow for neglecting decoherence due to gas scattering, the main decoherence mechanisms remaining are the scattering, absorption and the emission of black-body radiation. In the present paper, we are concerned with the environment temperature, i.e. with decoherence due to scattering and absorption of black-body radiation. In Fig.~\ref{fig:collapse_models}(left), we show the dependence of the interference visibility on the environment temperature. The figure shows predictions of quantum theory (solid lines) for a nanoparticle with a radius of $90\,$nm and a mass density of $5510\,\mathrm{kg/m^3}$ (Schott glass SF57HT). The dashed lines in the figure represent the predictions of the K\'arolyh\'azy model, and the shaded regions indicate where quantum theory predicts a higher interference visibility than the macrorealistic model. Note that the interference visibility predicted by the CSL model, the Di\'osi-Penrose model and the quantum-gravity model of Ellis is zero in all these plots. An observation of a non-zero interference visibility would, therefore, rule out all these models.
MAQRO will aim to measure the dependence of the interference visibility on various parameters like particle size and mass density in order to allow for a quantitative characterization of the decoherence mechanisms involved.

To keep the functional dependence in Fig.~\ref{fig:collapse_models}(left) simple, we assume that the internal temperature of the nanosphere is equal to the environment temperature. Of course, that is an idealized assumption because the nanoparticle will heat up while it is optically trapped. We show the functional dependence of the interference visibility on the nanoparticle's temperature in Fig.~\ref{fig:collapse_models}(right). Here, we assumed an environment temperature of $16.4\,$K, which is the result predicted by our thermal analysis for the spherical test volume shown in Fig.~\ref{fig:shield_geometry}. Note that our simulations predicting the interference visibility assumed an isotropic distribution of the black-body photons scattered and absorbed. Due to the shield geometry developed here it may be necessary to take into account an anisotropic distribution of the black-body radiation. This will be investigated in the future.

\section*{Conclusion}
Due to the inherent difficulties in combining the fundamental concepts of quantum theory and Einstein's theory of general relativity, it is often believed that the basic formulation of either theory may prove to be incomplete. Therefore, theoretical predictions are expected to deviate from sufficiently accurate measurements. Such deviations can be investigated in various ways: On the one hand, Einstein's equivalence principle as the foundation of general relativity can be put to the test. This has been done in numerous ground-based experiments and is proposed to be attempted at even higher accuracy for the future space missions ACES\cite{salomon2001cold}, MICROSCOPE\cite{Touboul2001a}, STEP\cite{sumner2007step}, and STE-QUEST\cite{schiller2010space}. On the other hand, the superposition principle as a central concept of quantum theory could be tested by addressing the question, whether quantum mechanics as we know it still holds for increasingly massive objects. Such an experiment is at the core of the proposed MAQRO mission\cite{kaltenbaek2012macroscopic}.

MAQRO addresses the question whether experiments to observe quantum superpositions of macroscopic objects could be successfully performed in space.
In order to test the coherence properties of such states against the currently proposed modifications of quantum theory, including those motivated by quantum gravity, it is essential to minimize the ``natural'' decoherence due to coupling with the environment which follows from conventional quantum mechanics. To perform such measurements, the quantum state must evolve freely for long periods of time in a cryogenic environment where black-body photon emission and scattering are suppressed to a high degree. 

For this purpose, a cryogenic instrument design based on passive cooling through radiative coupling to the cold void of space was developed. The concept does not employ cryogenic coolants, which not only make it cheaper and less complex but also avoids diffusive contamination. Optimal radiative shielding from the hot spacecraft surface, minimal conductive coupling to the spacecraft interior as well as an appropriate choice and placement of dissipative components are key requirements to achieve the lowest possible temperatures. We obtained a temperature of approximately 27 K for the optical bench and a temperature of 16 K for the test volume where experiments are performed. The latter value is limited by the imaging lens and can be improved by reducing the numerical aperture. We then showed that -- provided certain material properties are met for the test body -- the achievable temperatures allow testing the decoherence rates predicted by all major macrorealistic models, which seems intractable in ground-based experiments. Whilst the discussions in this paper focused on a specific type of experiment, the general design can be applied to other science experiments in space which aim for cryogenic temperatures of a compact optomechanical setup.

\section*{Acknowledgements}
RK acknowledges financial support from the Austrian Academy of Sciences (APART), the European Commission (Marie Curie) and the Austrian Research Promotion Agency (FFG) under project number 3589434. NK acknowledges support by the Alexander von Humboldt Stiftung. MA acknowledges support by the Austrian Science Fund FWF (SFB FoQuS), the European Research Council (ERC StG) and the European Commission (IP SIQS, ITN CQOM). Parts of the work presented in this paper were performed during study activities funded by the European Space Agency under contract Po P5401000400. We thank Eric Wille (ESA) for fruitful discussions.
\section*{References}
\bibliography{MAQRO_RS_references}
\bibliographystyle{unsrt}
\end{document}